\documentclass[aps,prl,showpacs,floatfix,twocolumn,superscriptaddress,longbibliography]{revtex4-2}

\usepackage{float}
\usepackage{color}
\usepackage{bm}
\usepackage{hyperref}
\usepackage{todonotes}
\usepackage{verbatim}
\usepackage{soul}
\usepackage{glossaries}
\usepackage{sidecap}

\def\Q{\ensuremath{\bm{Q}}}
\def\LNO{La$_{4}$Ni$_{3}$O$_{8}$}
\def\PNO{Pr$_{4}$Ni$_{3}$O$_{8}$}
\def\RNO{$R_4$Ni$_{3}$O$_{8}$}

\newacronym{BS}{BS}{bond-stretching}
\newacronym{RIXS}{RIXS}{Resonant Inelastic X-ray Scattering}
\newacronym{XAS}{XAS}{X-ray absorption spectrum}
\newacronym{EELS}{EELS}{electron energy loss spectroscopy}
\newacronym{LSCO}{LSCO}{La$_{2-x}$Sr$_{x}$CuO$_{4}$}
\newacronym{EPC}{EPC}{electron-phonon coupling}
\newacronym{CDW}{CDW}{charge density wave}
\newacronym{SDW}{SDW}{spin density wave}
\newacronym{FWHM}{FWHM}{full-width at half-maximum}
\newacronym{INS}{INS}{inelastic neutron scattering}
\newacronym{LNO}{LNO}{La$_{4}$Ni$_{3}$O$_{8}$}
\newacronym{PNO}{PNO}{Pr$_{4}$Ni$_{3}$O$_{8}$}
\newacronym{DFT}{DFT}{density functional theory}
\newacronym{GGA}{GGA}{generalized gradient approximation}

\begin{document}

\title{Strong Superexchange in a \texorpdfstring{$d^{9-\delta}$}{d9-d} Nickelate Revealed by Resonant Inelastic X-Ray Scattering} 

\author{J. Q. Lin}
\affiliation{Condensed Matter Physics and Materials Science Department, Brookhaven National Laboratory, Upton, New York 11973, USA}
\affiliation{School of Physical Science and Technology, ShanghaiTech University, Shanghai 201210, China}
\affiliation{Institute of Physics, Chinese Academy of Sciences, Beijing 100190, China}
\affiliation{University of Chinese Academy of Sciences, Beijing 100049, China}

\author{P. Villar Arribi}
\affiliation{Materials Science Division, Argonne National Laboratory, Lemont, Illinois 60439, USA}

\author{G. Fabbris}
\affiliation{Condensed Matter Physics and Materials Science Department, Brookhaven National Laboratory, Upton, New York 11973, USA}
\affiliation{Advanced Photon Source, Argonne National Laboratory, Lemont, Illinois 60439, USA}

\author{A. S. Botana}
\affiliation{Department of Physics, Arizona State University, Tempe, Arizona 85287, USA}

\author{D. Meyers}
\affiliation{Condensed Matter Physics and Materials Science Department, Brookhaven National Laboratory, Upton, New York 11973, USA}
\affiliation{Department of Physics, Oklahoma State University, Stillwater, Oklahoma 74078, USA}

\author{H. Miao}
\affiliation{Condensed Matter Physics and Materials Science Department, Brookhaven National Laboratory, Upton, New York 11973, USA}
\affiliation{Material Science and Technology Division, Oak Ridge National Laboratory, Oak Ridge, Tennessee 37830, USA}

\author{Y. Shen}
\affiliation{Condensed Matter Physics and Materials Science Department, Brookhaven National Laboratory, Upton, New York 11973, USA}

\author{D. G. Mazzone}
\affiliation{Condensed Matter Physics and Materials Science Department, Brookhaven National Laboratory, Upton, New York 11973, USA}
\affiliation{Laboratory for Neutron Scattering and Imaging, Paul Scherrer Institut, 5232 Villigen PSI, Switzerland}

\author{J. Feng}
\altaffiliation[Present address: ]{CAS Key Laboratory of Magnetic Materials and Devices, Ningbo Institute of Materials Technology and Engineering, Chinese Academy of Sciences, Ningbo 315201, China and Zhejiang Province Key Laboratory of Magnetic Materials and Application Technology, Ningbo Institute of Materials Technology and Engineering, Chinese Academy of Sciences, Ningbo 315201, China}
\affiliation{Sorbonne Universit\'{e}, CNRS, Laboratoire de Chimie Physique-Mati\`{e}re et Rayonnement,UMR 7614, 4 place Jussieu, 75252 Paris Cedex 05, France}
\author{S. G. Chiuzb\u{a}ian}
\affiliation{Sorbonne Universit\'{e}, CNRS, Laboratoire de Chimie Physique-Mati\`{e}re et Rayonnement,UMR 7614, 4 place Jussieu, 75252 Paris Cedex 05, France}
\affiliation{Synchrotron SOLEIL, L'Orme des Merisiers, Saint-Aubin, BP 48, 91192 Gif-sur-Yvette, France}

\author{A. Nag}
\author{A. C. Walters}
\author{M. Garc\'{i}a-Fern\'{a}ndez}
\author{Ke-Jin Zhou}
\affiliation{Diamond Light Source, Harwell Science and Innovation Campus, Didcot, Oxfordshire OX11 0DE, United Kingdom}

\author{J. Pelliciari}
\author{I. Jarrige}
\affiliation{National Synchrotron Light Source II, Brookhaven National Laboratory, Upton, NY 11973, USA}

\author{J. W. Freeland}
\affiliation{Advanced Photon Source, Argonne National Laboratory, Lemont, Illinois 60439, USA}
\author{Junjie Zhang}
\affiliation{Materials Science Division, Argonne National Laboratory, Lemont, Illinois 60439, USA}
\affiliation{Institute of Crystal Materials, Shandong University, Jinan, Shandong 250100, China}
\author{J. F. Mitchell}
\affiliation{Materials Science Division, Argonne National Laboratory, Lemont, Illinois 60439, USA}

\author{V. Bisogni}
\affiliation{National Synchrotron Light Source II, Brookhaven National Laboratory, Upton, NY 11973, USA}

\author{X. Liu}\email[]{liuxr@shanghaitech.edu.cn}
\affiliation{School of Physical Science and Technology, ShanghaiTech University, Shanghai 201210, China}

\author{M. R. Norman}\email[]{norman@anl.gov}
\affiliation{Materials Science Division, Argonne National Laboratory, Lemont, Illinois 60439, USA}

\author{M. P. M. Dean}\email[]{mdean@bnl.gov}
\affiliation{Condensed Matter Physics and Materials Science Department, Brookhaven National Laboratory, Upton, New York 11973, USA}

\date{\today}

\begin{abstract}
The discovery of superconductivity in a $d^{9-\delta}$ nickelate has inspired disparate theoretical perspectives regarding the essential physics of this class of materials. A key issue is the magnitude of the magnetic superexchange, which relates to whether cuprate-like high-temperature nickelate superconductivity could be realized. We address this question using Ni $L$-edge and O $K$-edge spectroscopy of the reduced $d^{9-1/3}$ trilayer nickelates \RNO{} (where $R$=La,Pr) and associated theoretical modeling. A magnon energy scale of $\sim80$~meV resulting from a nearest-neighbor magnetic exchange of $J = 69(4)$~meV is observed, proving that $d^{9-\delta}$ nickelates can host a large superexchange. This value, along with that of the Ni-O hybridization estimated from our O $K$-edge data, implies that trilayer nickelates represent an intermediate case between the infinite-layer nickelates and the cuprates. Layered nickelates thus provide a route to testing the relevance of superexchange to nickelate superconductivity.
\end{abstract}

\maketitle

\begin{figure}
\center
\includegraphics[width = 0.45\textwidth]{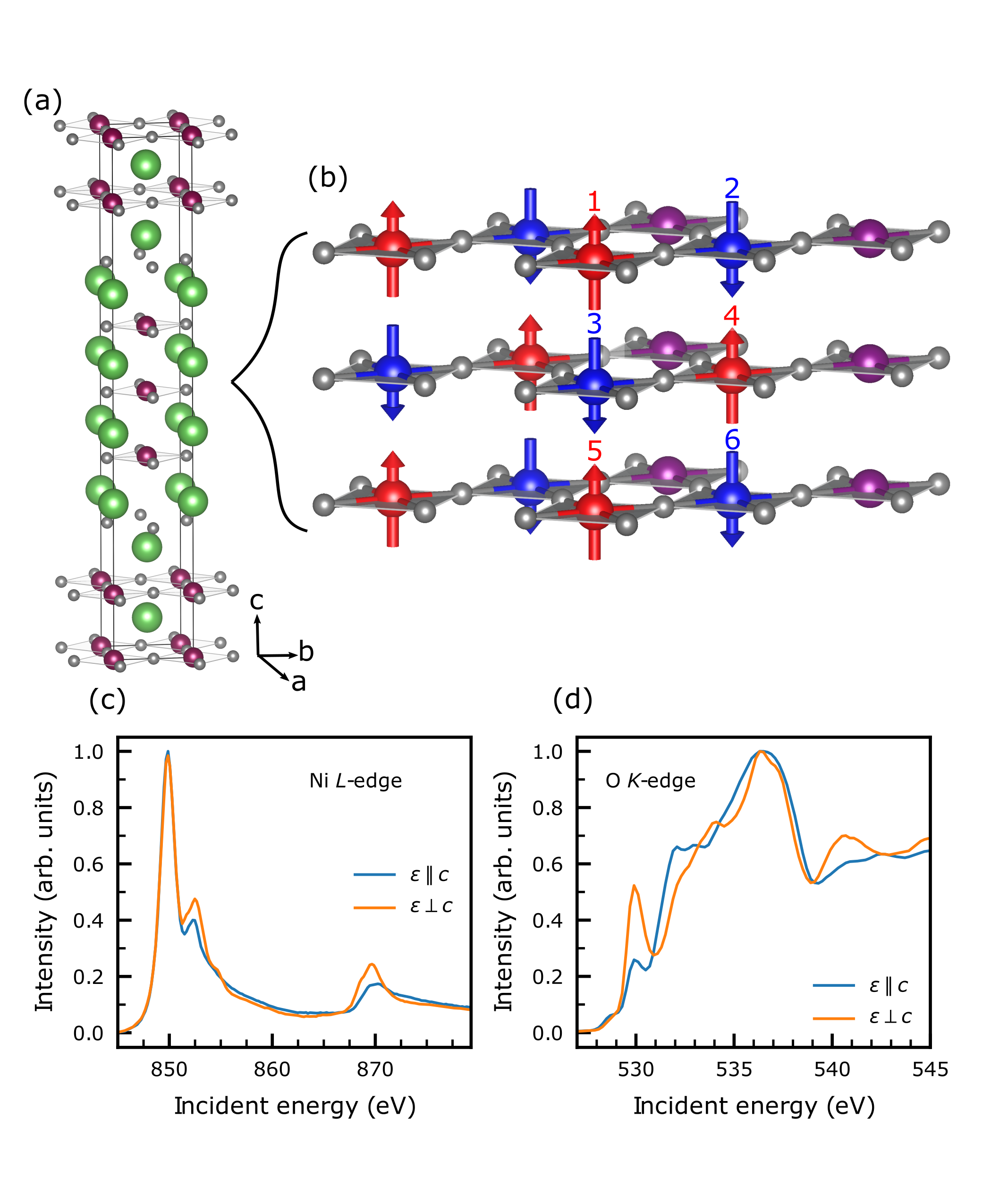}
\caption{Crystal structure and \acrfull*{XAS}. (a) Unit cell of \LNO{} and \PNO{} with Ni in purple, O in gray and La/Pr in green \cite{Momma2011VESTA}. (b) The active trilayer nickel-oxide planes in \LNO{} with an illustration of the diagonal stripe-ordered state \cite{Zhang2016stacked}. Ni sites with extra hole character (with respect to the $d^9$ magnetic rows) are in purple ($S=0$), whereas Ni up and down spins in the magnetic rows are colored red and blue, respectively ($S=1/2$). (c)\&(d) \gls*{XAS} data of \LNO{} measured in total fluorescence yield mode with polarization perpendicular and approximately parallel to the sample $c$-axis for (c) the Ni $L$-edge and (d) the O $K$-edge.}
\label{fig_xas_structure}
\end{figure}

Ever since the discovery of superconductivity in the cuprates \cite{Bednorz1986possible}, researchers have been searching for related unconventional high-temperature ($T_\text{c}$) superconductors based on different transition metal ions \cite{Norman2016materials, Adler2018correlated, Norman2020entering}. Nickel, given its proximity to copper in the periodic table, represents an obvious target element. A popular concept has been to try to realize materials with Ni$^{1+}$: $3d^9$ ions with planar oxygen coordination residing in layers, as it was conjectured that this would mimic the strong magnetic superexchange that was proposed to be important for cuprate superconductivity \cite{Anisimov1999electronic}. The appropriateness of this assumption in layered $R$NiO$_2$ materials ($R$= La, Pr, Nd) was, however, questioned as the predicted increase in charge-transfer energy in $R$NiO$_2$, with respect to cuprates, would be expected to reduce the superexchange \cite{Lee2004infinite}. Superconductivity at a relatively modest $T_\text{c}\approx 15$~K in Nd$_{1-x}$Sr$_x$NiO$_2$ was nonetheless reported \cite{Li2019superconductivity}. This has motivated many studies, often conflicting, concerning the nature of the normal-state electronic structure and correlations in these and related materials \cite{ Goodge2020doping, Osada2020superconducting, AriandoPRL, Hepting2020electronic, Bixia2020synthesis, Botana2020similarities, Sakakibara2019model, Jiang2019critical, Nomura2019formation, Wu2020robust, Zhang2020effective, Zhang2020self, Werner2020nickelate, Hu2019two, Zhang2020type, Liu2020electronic, Karp2020many, Been2020theory, Lang2020where, Wang2020hunds, Nomura2020magnetic,Kapeghian2020electronic,  Sawatzky2019superconductivity}. $R$NiO$_2$ materials are the infinite-layer members of a  series of low-valence (with d$^{9-\delta}$ filling)  layered nickelates $R_{n+1}$Ni$_n$O$_{2n+2}$ where $n$ represents the number of NiO$_2$ layers per formula unit  \cite{Crespin1983reduced, Hayward1999sodium, Hayward2003synthesis, Zhang2017large}. Given the important role of charge-transfer and superexchange in many theories of unconventional superconductivity, determining trends for these quantities is highly important for understanding nickelate superconductivity and potentially discovering new nickelate superconductors \cite{Anderson1987resonanting,Scalapino2012common}. Among the known members of this nickelate family, trilayer materials shown in Fig.~\ref{fig_xas_structure}(a) are ideal for testing the fundamental aspects of the analogy between layered nickelates and cuprates. This is because complications from rare-earth self-doping, $c$-axis coupling and inhomogeneous samples are less severe in \RNO{} than in $R$NiO$_2$ \cite{Poltavets2010bulk, Botana2016charge, Zhang2017large}.

\begin{figure*}
\center
\includegraphics[width=\linewidth]{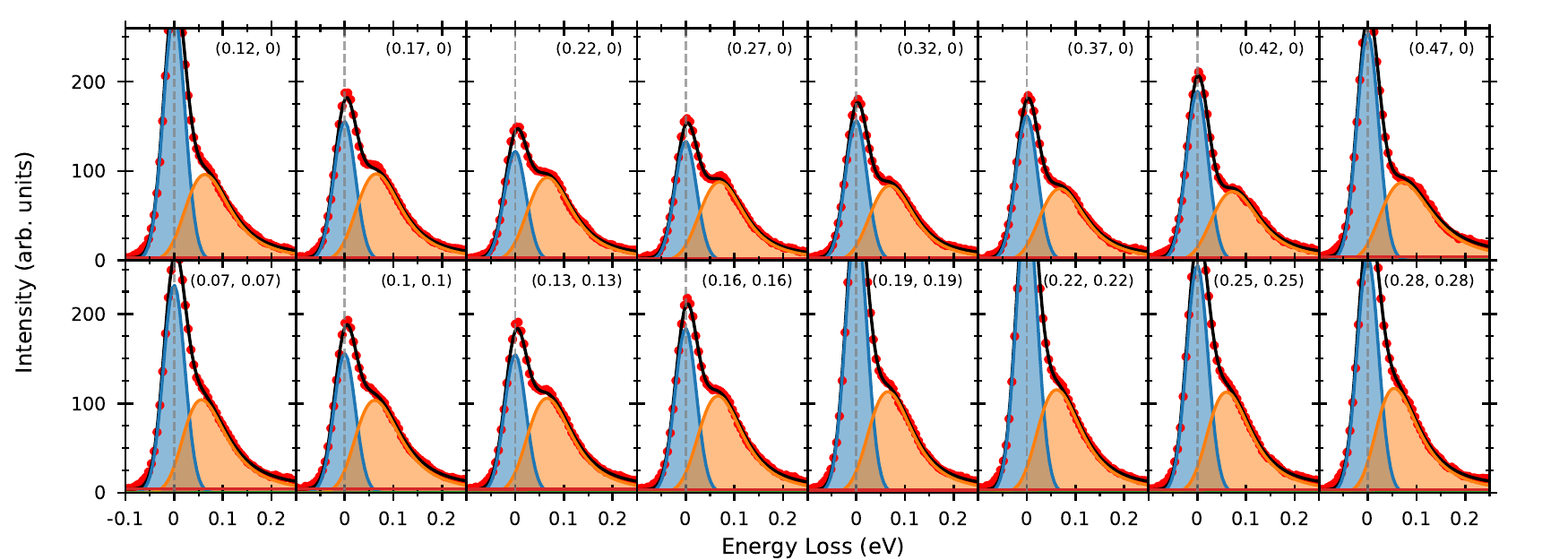}
\caption{RIXS spectra of \LNO{} as a function of \Q{} at the resonant energy of the magnon 852.7~eV \cite{supp}. Data are shown as red points and the fit is shown as a black line, which is composed of the magnetic excitation in orange and the elastic line in blue. The in-plane \Q{} of the measured spectrum is denoted in the top right of each panel.}
\label{fig_dispersion}
\end{figure*}

In this Letter, we combine \gls*{RIXS} with first principles calculations and theoretical modeling to characterize the magnetic exchange in trilayer \RNO{} that fall in the overdoped regime of cuprates in terms of electron count. We find a near-neighbor exchange of $J=69(4)$~meV, in good accord with our first principles calculations. This demonstrates that these reduced nickelates indeed have a strong superexchange (within a factor of two of the cuprates). By comparing the O $K$-edge pre-peak intensity to that of cuprates and infinite-layer nickelates, we argue that these trilayer materials are intermediate between cuprates and infinite-layer nickelates. Based on this, we suggest that electron-doping \RNO{} materials provides a compelling route to testing the relevance of superexchange for nickelate superconductivity.

\RNO{} ($R=$La, Pr) single crystals were prepared by synthesizing their parent Ruddlesden-Popper phases and reducing them in H$_2$/Ar gas as described previously \cite{Zhang2017large, supp}. The resulting samples are single-phase crystals with a tetragonal unit cell (I4/mmm space group) and lattice constants of $a=b=3.97$~\AA{}, $c=26.1$~\AA{}. The trilayer R$_4$Ni$_3$O$_8$ phase is shown in Fig.~\ref{fig_xas_structure}(a); panel (b) zooms on the Ni-O planes. These samples have an effective hole-doping of $\delta=1/3$. Reciprocal space is indexed in terms of scattering vector $\Q=(2\pi/a, 2\pi/a, 2\pi/c)$. Both La and Pr materials are rather similar regarding their high- and medium-energy physics such as spin states and orbital polarization \cite{Zhang2017large}. The primary difference is that \LNO{} (which exhibits strong antiferromagnetic spin fluctuations \cite{AppRoberts}) has stripe order than opens up a small insulating gap \cite{Zhang2017large}, whereas \PNO{} remains metallic without long-range order. Since the more ordered and insulating nature of \LNO{} compared to \PNO{} is expected to give sharper magnetic RIXS spectra, we focus on the former material for this paper.

We used \gls*{XAS} to confirm the expected electronic properties of the \LNO{} samples. The Ni $L$-edge spectrum from 846-878~eV is shown in Fig.~\ref{fig_xas_structure}(c). The strongest feature around 850~eV is the La $M_4$ edge, which is followed by the Ni $L_3$ and $L_2$ edges at 852 and 870~eV respectively. Substantial linear dichroism is apparent, especially at the $L_2$ edge where the spectrum is not obscured by the La $M_4$ edge, indicating that the unoccupied $3d$ states are primarily $x^2-y^2$ in character \cite{Zhang2017large}. The overall spectral shape is very similar to that seen in cuprates \cite{Chen1991electronic, Chen1992out, Brookes2015stability}, consistent with a $d^9\underline{L}$ configuration, with no indication for a high-spin $d^8$ component of the holes \cite{Zhang2017large}. This is reasonable, since the planar coordination of Ni leads to a large splitting between the $x^2-y^2$ and $3z^2-r^2$ states, which is expected to out-compete the Hunds exchange coupling, thus favoring a low-spin ground state \cite{Fabbris2016orbital}. The O $K$-edge spectrum around 525-545~eV in Fig.~\ref{fig_xas_structure} shows a pre-peak feature around 532~eV, which is known to indicate hybridization between the Ni $3d$ and O $2p$ states \cite{Chen1991electronic, Chen1992out, Zhang2017large}. Our measurements find that this pre-peak has a strong linear dichroism as well, as observed in cuprates \cite{Chen1992out}.

We then performed \gls*{RIXS} to study the low-energy degrees of freedom. High energy-resolution \gls*{RIXS} measurements were performed at I21 at the Diamond Light Source with a resolution of 45~meV and at NSLS-II with a resolution of 30~meV. All RIXS data shown were taken at a temperature of 20~K using a fixed horizontal scattering angle of $2\theta=154^{\circ}$ and x-ray polarization within the horizontal scattering plane ($\pi$ polarization). Different momenta were accessed by rotating the sample about the vertical axis, such that the projection of the scattering vector varies. $(H, 0)$ and $(H, H)$ scattering planes were accessed by rotating the sample about its azimuthal angle. Figure~\ref{fig_dispersion} plots low-energy \gls*{RIXS} spectra of \LNO{} as a function of \Q{}. A relatively strong elastic line is present for all \Q{} likely arising from apical oxygen removal during sample preparation, which induces internal strain in the samples. In the 70-90~meV energy range, a weakly dispersive, damped feature is observed.  Based on the energy of the feature, this could either be magnetic or the bond-stretching phonon common to complex oxides \cite{Fabbris2017doping, Betto2017three, Nag2020many, Lin2020strongly, Devereaux2016directly}. It is known, however, that the intensity of the bond-stretching phonon increases like $|\Q|^2$, inconsistent with what we find \cite{Devereaux2016directly, Lin2020strongly} (see Fig.~\ref{fig_dispersion}).  The peak also resonates slightly above the Ni $L$ edge \cite{supp}, which is also consistent with a magnetic origin \cite{Fabbris2017doping}. On the basis of these observations, we assign this feature to magnetic excitations. Further supporting this assignment we note that \PNO{} spectra exhibit a weaker, more damped paramagnon excitation, which is expected as this compound is metallic with spin-glass behavior \cite{supp, Huangfu2020shortrange}. Below, we will demonstrate a consistency between this mode and analytical modeling and \gls*{DFT}. 

\begin{figure}
\center
\includegraphics[width=0.4\textwidth]{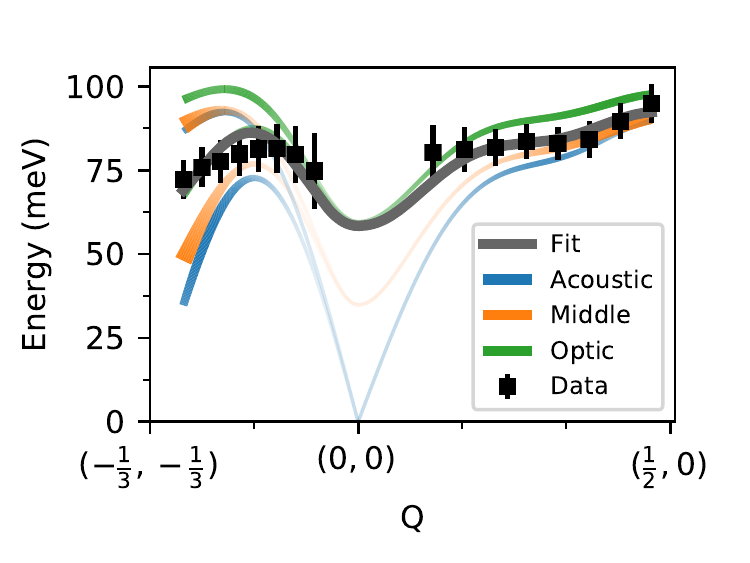}
\caption{Magnetic dispersion of \LNO{}. Black points are the extracted energies of the magnetic excitation. The gray line is the fit to the experimental dispersion, which is composed of the weighted sum of three dispersive magnons, called the acoustic, middle and optic modes, which are plotted as blue, orange and green lines, respectively. The thickness of all three lines represents the predicted intensity of the modes \cite{supp}. The doubling of the modes from $(-\frac{1}{3}, -\frac{1}{3})$ to $(0, 0)$ arises from magnetic twinning \cite{supp}.}
\label{fig_fit}
\end{figure}

In order to analyze the magnetic dispersion, we fit the low-energy \gls*{RIXS} spectra with the sum of a zero-energy Gaussian fixed to the experimental energy resolution in order to account for the elastic line and a damped harmonic oscillator to capture the magnetic excitation \cite{supp}. To model the magnetic interactions, we expect a leading contribution from the nearest-neighbor in-plane Ni-O-Ni superexchange, $J$. We further know that \LNO{}, like some other nickelates and cuprates, has a striped ground state with both spin and charge character, with an in-plane wavevector of $\Q = (1/3, 1/3)$ \cite{Zhang2016stacked, Zhang2019stripe}. This diagonal stripe order is illustrated in Fig.~\ref{fig_xas_structure}(b) \cite{supp}.  In each plane, we have two antiparallel spin rows (corresponding to $d^9$) separated by an anti-phase domain wall (corresponding to non-magnetic $d^8$).  This gives rise to six spins in the magnetic unit cell in a given trilayer, which we label as spins 1-6.  Nearest-neighbor spins within the planes in a given stripe are coupled by the superexchange $J$, which we expect to be the strongest interaction.  The antiphase domain wall is due to coupling between the magnetic stripes.  There are two potential couplings (super-superexchange), but we only expect one of them (the one along the tetragonal axes) to be significant, as the other involves a 90 degree pathway \footnote{Diagrams for the exchange pathways are shown in \cite{supp}}. As the planes are antiferromagnetically coupled \cite{Zhang2019stripe}, this gives rise to a positive $J_z$ coupling between successive layers (there is no evidence for magnetic coupling between the trilayers, so our model deals with only a single trilayer). We solved the resulting Heisenberg model in the spin-wave approximation \cite{[{Our model is essentially a trilayer generalization of }] Carlson2004spin}, which yields three dispersive modes (split by $J_z$), which we term the acoustic, middle and optic modes \cite{supp}. The energy of each of these three modes changes with in-plane momentum and the relative intensity of the modes is modulated by the out-of-plane momentum, which varies with in-plane momentum due to our fixed-scattering-angle configuration. From cuprates, we anticipate that the interlayer coupling will be of order 10~meV, below our energy resolution \cite{Reznik1996direct, Dean2014itinerant}. On this basis, we  analyzed our data in terms of the sum of the three magnon modes. The \gls*{RIXS} intensity for a particular acoustic, middle or optic magnetic mode $n$ in the $\pi-\sigma$ polarization channel is given by \cite{Haverkort2010theory}
\begin{equation}
 I_n(\bm{Q})=\left|\sum_i\bm{k}_{in}\cdot\bm{M}_{n,\bm{Q}}(\bm{r}_i)\right|^2
 \label{eq:intensity_1}
\end{equation}
where $\bm{k}_{in}$ is the incident wavevector and $\bm{M}_{n,\bm{Q}}(\bm{r}_i)$ is magnetization vector at site $i$ (i.e., the eigenvector of the $n^{th}$ spin-wave mode at $\bm{Q}$). This vector is in-plane since the ground-state moments are along $c$ \cite{Zhang2019stripe}. The final element of our model is to sum over the two tetragonal domains given the known magnetic twinning in \LNO{}: $(H, K)\rightarrow ({H, -K})$ \cite{Zhang2016stacked, Zhang2019stripe}. We determined the energies and eigenvectors of these modes from the resulting 12 by 12 secular matrix \cite{supp}, and computed the weighted sum of the three modes at each $\bm{Q}$. \footnote{Our spin-wave approach does not include any quantum renormalization factor, $Z_c$, of the spin wave energy, and this difference should be accounted for when comparing to cuprates when $Z_c$ has been accounted for, which requires exchange values $Z_c \sim 1.18$ times larger to reproduce the same magnon energy \cite{Singh1989thermodynamic, Peng2017influence}.} To estimate the magnetic exchange parameters in \LNO{}, we computed the energy of four different spin configurations in the above-mentioned magnetic cell \cite{supp}, and then mapped these energies to a Heisenberg model. This was done using \gls*{DFT} in the \gls*{GGA} approximation as implemented in the WIEN2k code \cite{Blaha2020WIEN2k1, Blaha2020WIEN2k2}. The experimentally determined insulating charge and spin stripe-ordered ground state (Fig.~1b) is obtained even at the \gls*{GGA} level given that the exchange splitting is larger than the bandwidth in this state. Adding a $U$ simply increases the size of the gap with respect to the GGA solution, but the nature of the ground state remains the same \cite{Poltavets2010bulk}. Results presented here are for \gls*{GGA}, but GGA+U results are presented in \cite{supp}. This yields $J= 71$~meV, $J_z= 13.6$~meV, and $J_1= 10.6$~meV. We fix $J_z= 13.6$~meV in our model since, because of our resolution and contamination from the elastic line, we cannot accurately estimate it from experiment. We then vary $J$ and $J_1$ to get the best fit. This fit yields $J=69(4)$~meV and $J_1=17(4)$~meV in good agreement with experiment, although the small difference in $J$ (2~meV) is likely coincidental. These exchange values can be rationalized from the single-layer analytic relation (i.e., ignoring $J_z$) that $E_\text{mag} \sim 4S\sqrt{J J_1}$ where $E_\text{mag}$ is the zone-boundary magnon energy. We overplot the magnetic dispersion with our theory analysis in Fig.~\ref{fig_fit} showing a good level of agreement. The model also captures the observed softening that occurs as \Q{} approaches $(-\frac{1}{3}, -\frac{1}{3})$. We also measured \PNO{} \cite{supp}, which is similar to \LNO{}, but metallic rather than insulating; the results show a lower-intensity damped magnetic excitation, which is expected in view of its metallicity \cite{Zhang2017large} and the spin-glass behavior reported for this material \cite{Huangfu2020shortrange}. The paramagnon energy in \PNO{} is  only slightly reduced compared to \LNO{}. Again, this is similar to cuprates where magnon-like excitations are seen for paramagnetic dopings \cite{Dean2013persistence}. 

%\todo[inline]{info on width of excitation}

Our rather large value of $J=69(4)$~meV is the principal result of this Letter. This magnetic exchange is 2.5 times larger than that of the $1/3$ doped nickelate La$_{2-x}$Sr$_x$NiO$_4$ which also has a diagonal stripe state (with $S=1$ $d^8$ magnetic rows and $S=1/2$ $d^7$ domain walls), though the two have comparable $J_1$ \cite{Boothroyd2003spin, Woo2005mapping, Fabbris2017doping}. $J$ for \LNO{} is, in fact, within a factor of two of cuprates, which have among the largest superexchange of any known material \cite{Coldea20110spin, LeTacon2011intense, Dean2013persistence, Peng2017influence}. This suggests, along with our \gls*{XAS} results, that these nickelates are strongly correlated charge-transfer materials. Two questions are apparent: Why is the superexchange in \LNO{} so large? And what is the relationship between trilayer nickelates and their infinite-layer counterparts?

\begin{figure}
\center
\includegraphics[width = 0.45\textwidth]{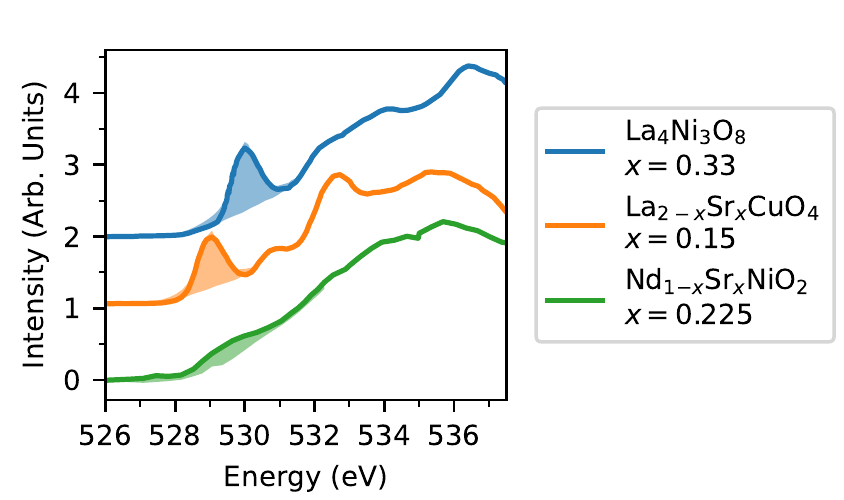}
\caption{Comparison of the O $K$-edge in-plane polarized pre-peak intensity, indicative of oxygen-hybridized holes, between different $d^{9-\delta}$ materials. Solid lines are \gls*{XAS} or \gls*{EELS}. The data and background (lineshape excluding the pre-peak) for Nd$_{1-x}$Sr$_x$NiO$_2$ are from Ref.~\cite{Goodge2020doping} and the data for La$_{2-x}$Sr$_x$CuO$_4$ are from Ref.~\cite{Chen1992out}. Further details are provided in \cite{supp}.}
\label{fig_prepeak}
\end{figure}

Given the 180 degree Ni-O-Ni bonds in the $d^9$ nickelates, superexchange is the most likely mechanism for generating their exchange interactions, as in the cuprates. In the charge-transfer limit, the strength of this interaction scales as $t^{4}_{pd}/\Delta^3$ where $t_{pd}$ is the hopping between the transition metal $x^2-y^2$ and oxygen $p\sigma$ orbitals, and $\Delta \equiv E_d - E_p$ is the energy difference between them. Large $p-d$ hopping and a small $\Delta$ implies a large ligand-hole character for the doped holes, as this is controlled  by the ratio $t_{pd}/\Delta$. We therefore fit the O $K$ pre-peak intensity to compare to the literature for La$_{2-x}$Sr$_x$CuO$_4$ \cite{Chen1991electronic} and Nd$_{1-x}$Sr$_x$NiO$_2$ \cite{Goodge2020doping} and show the results in Fig.~\ref{fig_prepeak}. The pre-peak in Nd$_{1-x}$Sr$_x$NiO$_2$ is significantly less prominent than in \LNO{} and La$_{2-x}$Sr$_x$CuO$_4$, but this appears to be primarily due to a broadened pre-peak rather than a lower integrated spectral weight, the broadening perhaps due to a spatially varying doping. The relative integrated weight per doped hole is 1.00(2):1.74(5):1.05(10) for  \LNO{}:La$_{2-x}$Sr$_x$CuO$_4$:Nd$_{1-x}$Sr$_x$NiO$_2$ and the equivalent ratios for the maximum pre-peak intensities are 1.00(4):1.85(9):0.37(6). The quoted errors are the uncertainty from the least-squares fitting algorithm. The largest systematic error likely arises from the doping inhomogeneity in the data from \cite{Goodge2020doping}. Thus \LNO{} has somewhat less admixture than La$_{2-x}$Sr$_x$CuO$_4$, but the difference is not enough to expect qualitatively different physics. The ratio we determine is in good accord with our observed magnetic exchange in \LNO{} being around half that of the cuprates \cite{LeTacon2011intense, Dean2012spin, Dean2013persistence, Peng2017influence}. This difference likely comes from $\Delta$ being larger in \LNO{} compared to La$_{2-x}$Sr$_x$CuO$_4$ ($t_{pd}$ is comparable in the three materials) \cite{Nica2020theoretical, McMahan1990}. The situation in NdNiO$_2$ is less certain given the large difference of the ratios (i.e., whether one considers the maximum intensity or the integrated weight). This will likely only be solved when higher-quality more homogeneous Nd$_{1-x}$Sr$_x$NiO$_2$ samples are prepared and studied in detail.  Still, it seems likely that the superexchange in NdNiO$_2$ is smaller, but still large enough that its potential contribution to superconductivity deserves serious consideration. Recent Raman scattering measurements estimate $J\approx 25$~meV \cite{Fu2019core},  which is consistent with this and  which likely arises from $\Delta$ being larger, though the enhanced $c$-axis coupling and the screening from the $R$ $5d$ states, which are predicted to be partially occupied NdNiO$_2$, could be playing a role as well.

In conclusion, we report the presence of a large superexchange $J=69(4)$~meV in \LNO{} -- the first direct measurement of superexchange in a $d^{9-
\delta}$ nickelate. This superexchange value is within a factor of two of values found in the cuprates, and this, coupled with a substantial O $K$ pre-peak, establishes the charge-transfer nature of this $d^{9-\delta}$ nickelate with substantial $d$-$p$ mixing. By comparing the O $K$-edge \gls*{XAS} spectra of \LNO{} to that of the cuprates and the infinite-layer nickelate, we establish that trilayer nickelates represent a case that is intermediate between them. This result is interesting in view of the widespread belief that increasing magnetic superexchange might promote higher-$T_c$ superconductivity   \cite{Zhang2017large, Botana2017electron, Norman2020entering}. Studying a series of layered nickelates would also provide a route to testing the relevance of superexchange to nickelate superconductivity given the variation in their nominal Ni valence.

\begin{acknowledgments}
Work at Brookhaven National Laboratory was supported by the U.S.\ Department of Energy, Office of Science, Office of Basic Energy Sciences. Work at Argonne was supported by the U.S.\ Department of Energy, Office of Science, Basic Energy Sciences, Materials Science and Engineering Division (Materials Science Division) and Scientific User Facilities Division (Advanced Photon Source). X. L.\ and J.Q.L.\ were supported by the ShanghaiTech University startup fund, MOST of China under Grant No.~2016YFA0401000, NSFC under Grant No.~11934017 and the Chinese Academy of Sciences under Grant No.~112111KYSB20170059.  Work at the Advanced Photon Source was supported under Contract No. DE-AC02-06CH11357. A.B.\ acknowledges the support from NSF DMR 2045826. We acknowledge Diamond Light Source for time on Beamline I21 under Proposal 22261 producing the data shown in Fig.~2. This research used resources at the SIX beamline of the National Synchrotron Light Source II, a U.S.\ Department of Energy (DOE) Office of Science User Facility operated for the DOE Office of Science by Brookhaven National Laboratory under Contract No.~DE-SC0012704. We acknowledge Synchrotron SOLEIL for provision of synchrotron radiation facilities at the SEXTANTS beamline.
\end{acknowledgments}

\bibliography{refs}
\end{document}

\title{Supplemental Material: Strong Superexchange in a \texorpdfstring{$d^{9-\delta}$}{d9-d} Nickelate Revealed by Resonant Inelastic X-Ray Scattering}

\date{\today}

\maketitle

\renewcommand{\thefigure}{S\arabic{figure}}

\section{Further details of the single-crystal synthesis}
Single-crystal growth of $R_4$Ni$_3$O$_{10}$ ($R = $La, Pr) was performed as described Refs.~\cite{Zhang2016stacked, Zhang2017large}. The parent Ruddlesden-Popper phases were prepared in a  floating zone furnace (HKZ-1, SciDre GmbH) with 20 bar O$_2$ for $R=$La and 140~bar for $R$=Pr. Oxygen was flowed at a rate of 0.1~l/min during growth and the feed and seed rods were counter-rotated at 30~r.p.m.\ and 27~r.p.m., respectively, to improve zone homogeneity. The traveling speed of the seed was 4~mm/h and the growth time was 30 hours. 438-phase crystals were obtained by reducing the 4310-specimens in 4 mol \% H$_2$/Ar gas at 350$^{\circ}$C for five days. The resulting samples have appreciable residual strain and are very brittle, so they were mounted on copper plates for transport. 

\section{Resonant behavior}
\begin{figure}
\center
\includegraphics[width = 0.5\textwidth]{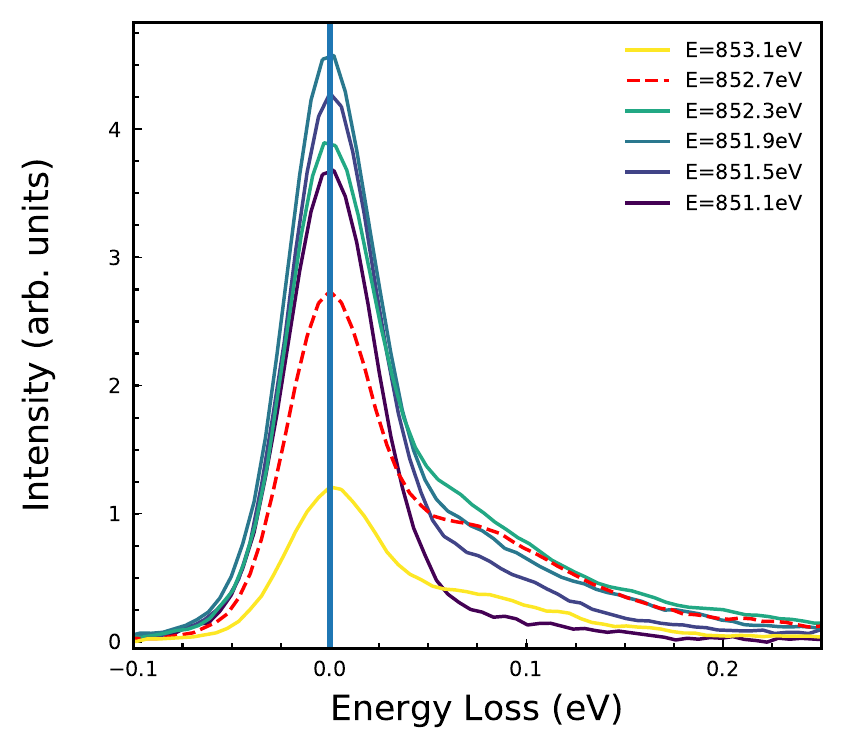}
\caption{RIXS spectra of \LNO{} at $(-0.44, 0)$ as a function of incident energy around the Ni $L_3$ edge. The chosen working energy that optimizes the magnon intensity, 852.7~eV, is highlighted using a dashed line and occurs above the maximum in the elastic line resonance, further confirming the magnetic nature of the observed inelastic excitations \cite{Fabbris2017doping}. }
\label{fig_Edep}
\end{figure}

In Fig.~\ref{fig_Edep} we show the resonant behavior of the magnon. The magnon peak is visible at several energies from 851.5 to 853.1~eV exhibiting a Raman-like behavior in which it appears at a constant energy loss, rather than a constant final x-ray energy. We find that the magnon is strongest at 852.7~eV (as emphasized by the dashed line at this energy). This is well above the La $M_4$-edge at 849~eV, further confirming the magnetic origin of the magnon.

\section{Fitting of the RIXS data}
The spectra were fitted with a Gaussian function for the elastic peak, a damped harmonic oscillator model for the paramagnon and a quadratic background. The inelastic peak was convoluted with a Gaussian function to account for the energy resolution. This lineshape is described by nine parameters, but only six parameters are free to vary in the fit. For the Gaussian lineshape describing the elastic peak, the center and the width are fixed by measurements of a graphite elastic reference sample, only the amplitude is free to vary. For the inelastic mode, the temperature is fixed, and the center, width and amplitude are free. For the quadratic background, we use a function $f(x) = b$ for $x<0$ and $f(x) = a*x^{2}+ b$ for $x>0$, $a$ and $b$ are free parameters. Prior to computing the final fit, we performed an initial fit in which the elastic energy was allowed to vary, which we used to shift the spectra in energy such that the elastic energy is set to exactly zero. 

\section{Comparison of L\tlc{a}$_4$N\tlc{i}$_3$O$_8$ and P\tlc{r}$_4$N\tlc{i}$_3$O$_8$}
\begin{figure*}
\center
\includegraphics[width=\linewidth]{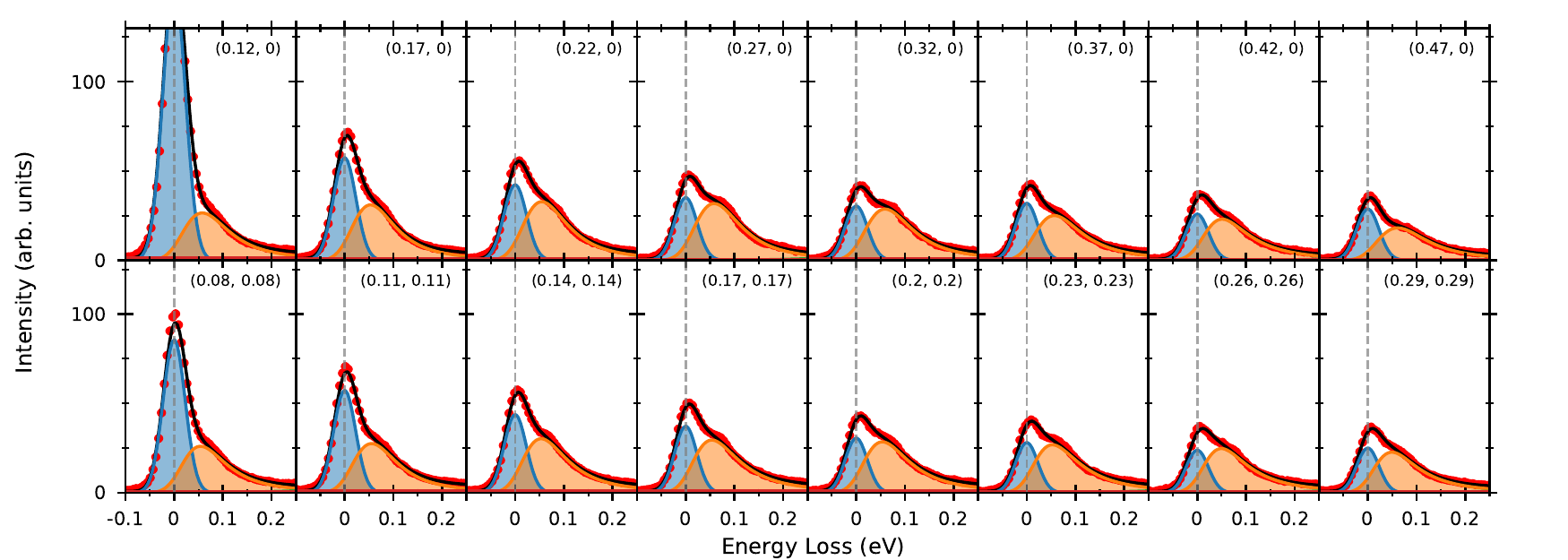}
\caption{RIXS spectra of \PNO{} as a function of \Q{} at the resonant energy of the magnon 852.7~eV. Data are shown as red points and the fit is shown as a black line, which is composed of the magnetic excitation in orange and the elastic line in blue. The in-plane \Q{} of the measured spectrum is denoted in the top right of each panel. Note that the scale of the y-axis is half of that in Fig.~2 of the main text.}
\label{fig_dispersion_SM}
\end{figure*}

\begin{figure*}
\center
\includegraphics[width=0.5\textwidth]{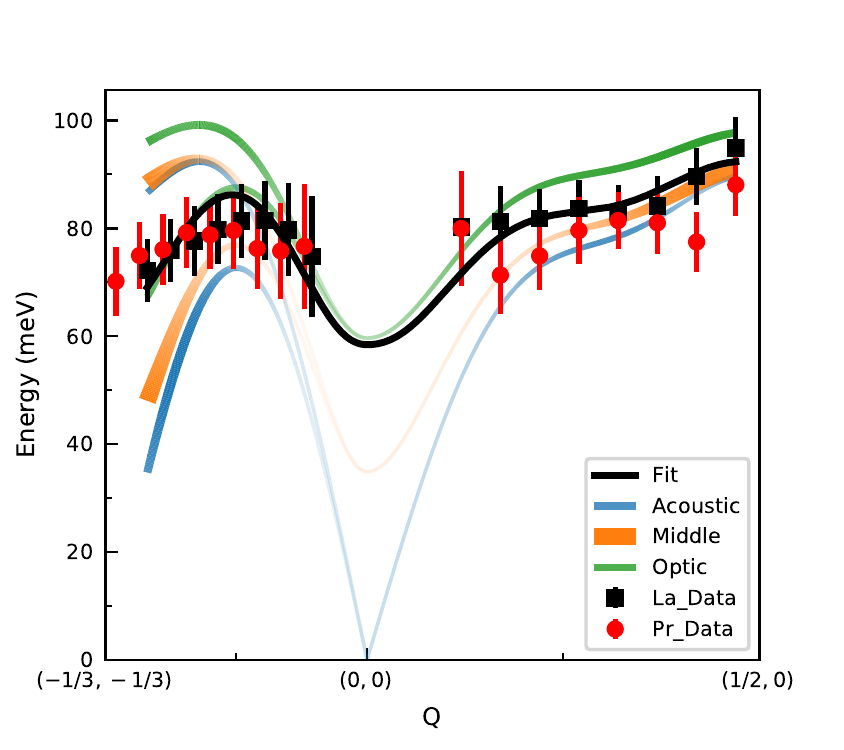}
  \caption{Magnetic dispersion of \LNO{} and \PNO{}. This figure is the same as Fig.~3 of the main text, but with data for \PNO{} added. Black/red points are the extracted energies of the magnetic excitation for \LNO{}/\PNO{}. The black line is the fit to the experimental dispersion of \LNO{}, which is composed of the weighted sum of three dispersive magnons, called the acoustic, middle and optic modes, which are plotted as blue, orange and green lines, respectively.}
  \label{fig_fit_Pr}
\end{figure*}

The difference between \LNO{} and \PNO{} has been studied in prior x-ray absorption and \gls*{DFT}-based work \cite{Zhang2017large}. This study concluded that both materials are rather similar regarding their high- and medium-energy physics such as spin states, orbital polarization, etc. The primary difference is that stripe order opens a small insulating gap in \LNO{}, whereas \PNO{} remains metallic without long-range order. \PNO{} was later reported to have spin-glass behavior likely coming from short-range stripe correlations \cite{Huangfu2020shortrange}. Since the more ordered and insulating nature of \LNO{} compared to \PNO{} is expected to give sharper magnetic RIXS spectra, we focused on the former material for this paper, but we also took data on \PNO{} as shown in Fig.~~\ref{fig_dispersion_SM}. A similar energy peak is observed which is slightly broadened and less intense compared to \LNO{}. Fitting the spectra for \PNO{} in the same way as was done for \LNO{} yields a value of the near-neighbor exchange perhaps 10\% lower, but overall the two materials are very similar (Fig.~\ref{fig_fit_Pr}). 

\section{Theory of magnetic excitations in the stripe-ordered state}
\begin{figure}
\center
\includegraphics{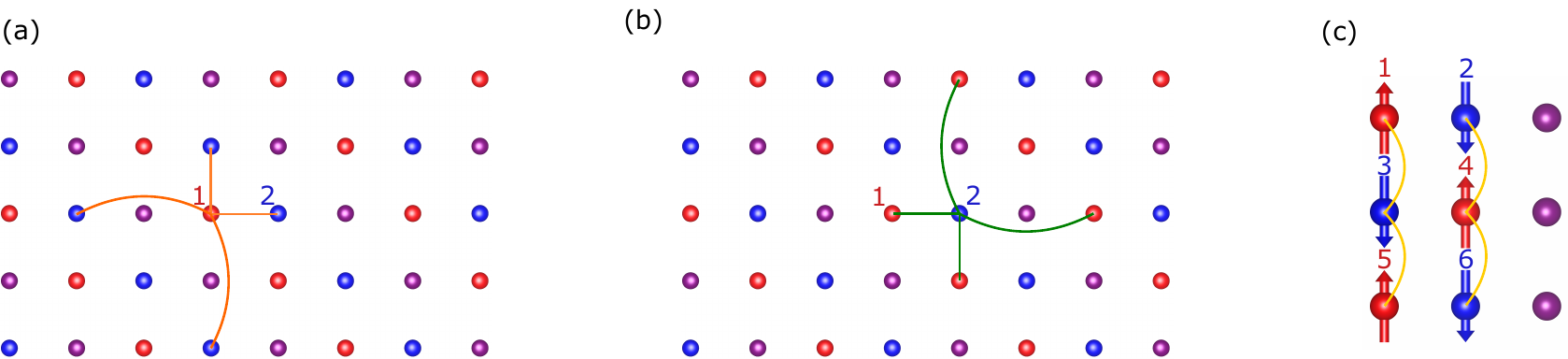}
\caption{Coupling of the spins in the stripe-ordered state for: (a) a spin up in the top layer of the trilayer, (b) a spin down in the top layer, and (c) between the different layers. For simplicity, only the Ni sites are shown with red, blue and purple denoting up-spin, down-spin and hole states as in Fig.~1 of the main text.}
\label{fig_couplings}
\end{figure}

In this Section, we compute the dispersion relation and RIXS intensity for the magnons in a diagonal stripe state. To reproduce the stripe order shown in Fig.~1(b) of the main text, we consider a model with the following interactions illustrated in Fig.~\ref{fig_couplings}. $J$ couples nearest-neighbor spins within the same stripe, $J_1$ couples spins across the stripes in the $[1,0,0]$ direction, and $J_z$ couples spins between layers within the trilayer in the $[0,0,1]$ direction. A further $J_2$ coupling across the stripes along the $[1,1,0]$ direction was also considered, but its effect could not be distinguished in the measured RIXS spectra, so it was omitted. This is expected as this super-superexchange contribution is weak given the 90 degree Ni-Ni-Ni pathway that is involved. We also ignore any single-ion anisotropy, again because it would be difficult to detect given the width of the elastic line.  The in-plane lattice vectors for the structural unit cell are
\begin{equation}
\begin{aligned}
 \bm{a_1}=(a,0,0),&
 &\bm{a_2}=(0,a,0)
% &\bm{a_3}=(0,0,c),
 \label{eq:1_latt_vec}
\end{aligned}
\end{equation}
and for the magnetic unit cell are
\begin{equation}
\begin{aligned}
 \bm{a_1^\text{mag}}=(3a,0,0),&
 &\bm{a_2^\text{mag}}=(-a,a,0).
% &\bm{a_3^\text{mag}}=(a/2,a/2,c/2).
 \label{eq:1_mag_vec}
\end{aligned}
\end{equation}
where, for simplicity, we have assumed a primitive magnetic cell with non-orthogonal lattice vectors.  Along the $c$ direction, we have trilayers separated by the body-centered translation ($a$/2, $a$/2, $c$/2).  Since there are no observable correlations between trilayers \cite{Zhang2019stripe}, we consider only a single trilayer here.  The layers within a trilayer are in registry along $c$, with each layer separated by the  interlayer distance $d\approx c$/8.  The scattering vector $\bm{Q}$ is presented in normalized units with $a=c=1$. 

\subsection{Dispersion relation}
We proceed to calculate the magnon dispersion for the diagonal stripe state by generalizing the torque equation formalism of Carlson {\it et al.}~\cite{Carlson2004spin} to the trilayer case (equivalent results can be obtained using the less transparent Holstein-Primakoff treatment \cite{Carlson2004spin}). According to neutron scattering data, the spins in the ground state are oriented along $c$ \cite{Zhang2019stripe}.  Therefore, the generalized torque equations for the spins reduce to:
\begin{equation}
\begin{aligned}
  &\frac{dS^x_{\bm{r},i}}{dt} = -\frac{1}{\hbar}\left( S^y_{\bm{r},i} \sum_{\bm{r'},j} J^{ij}_{\bm{r}\bm{r'}} S^z_{\bm{r}',j} - S^z_{\bm{r},i} \sum_{\bm{r'},j} J^{ij}_{\bm{r}\bm{r'}} S^y_{\bm{r}',j} \right),\\
  &\frac{dS^y_{\bm{r},i}}{dt} = -\frac{1}{\hbar}\left( S^z_{\bm{r},i} \sum_{\bm{r'},j} J^{ij}_{\bm{r}\bm{r'}} S^x_{\bm{r}',j} - S^x_{\bm{r},i} \sum_{\bm{r'},j} J^{ij}_{\bm{r}\bm{r'}} S^z_{\bm{r}',j} \right),\\
  &\frac{dS^z_{\bm{r},i}}{dt} \approx 0,
 \label{eq:2_torque_eqs}
\end{aligned}
\end{equation}
where $\bm{r},\bm{r'}$ label the positions of the spins in different magnetic unit cells and the indices $i,j$ label the spins within each magnetic unit cell $(i,j=1,...,6)$ as shown in Fig.~S2. We seek sinusoidal solutions of the form
\begin{equation}
 S^x_{\bm{r},i}=S^x_{i}  \exp[i(\bm{Q}\cdot\bm{r}-\omega t)]\ , \ \ \  S^y_{\bm{r},i}=S^y_{i}  \exp[i(\bm{Q}\cdot\bm{r}-\omega t)], 
 \label{eq:3_solutions}
\end{equation}
and we set $S^z_{\bm{r},i}=\pm S$ with the sign given by the orientation of the spin in the ground state. To start, we identify the couplings $J^{ij}_{\bm{r}\bm{r'}}$ that connect the spins at different lattice positions, with the origin taken to be the location of spin 1. We distinguish two groups of spins, ($S_1$, $S_3$, $S_5$) and ($S_2$, $S_4$, $S_6$), with each group having equivalent in-plane locations due to the $c$-axis translational symmetry. Their couplings are
\begin{itemize}
 \item $S_1$ ($\bm{r}=\bm{0}$) couples to $S_2$ twice with $J$ ($\bm{r}^\prime = \bm{a_1}$, $\bm{r}^\prime = \bm{a_2}$) and twice with $J_1$ ($\bm{r}^\prime=-2\bm{a_1}$, $\bm{r}^\prime=-2\bm{a_2}$). The same applies for $S_3$ ($S_5$) coupled to $S_4$ ($S_6$).
 \item $S_2$ ($\bm{r}=\bm{a_1}$) couples to $S_1$ twice with $J$ ($\bm{r}'=\bm{0}$, $\bm{r}'=\bm{a_1}-\bm{a_2}$) and twice with $J_1$ ($\bm{r}^\prime=3\bm{a_1}$, $\bm{r}^\prime=\bm{a_1}+2\bm{a_2}$). The same applies for $S_4$ ($S_6$) coupled to $S_3$ ($S_1$).
\end{itemize}
For the couplings along [0,0,1], we have:
\begin{itemize}
 \item $S_1$ ($S_2$) couples to $S_3$ ($S_4$) with $J_z$.
 \item $S_5$ ($S_6$) couples to $S_3$ ($S_4$) with $J_z$.
 \item $S_3$ ($S_4$) couples with $J_z$ to $S_1$ ($S_2$) and to $S_5$ ($S_6$).
\end{itemize}
With this information, we can write the torque equations for each of the six spins in the magnetic unit cell, for example:
\begin{equation}
\begin{aligned}
   \frac{dS^x_{\bm{0},1}}{dt}&=-\frac{1}{\hbar}  \Biggl\{ S^y_{\bm{0},1}(-S)\Big(2J+2J_1+J_z\Big) \\
  & -S\Biggl[J \Big(S^y_{\bm{a_1},2}+S^y_{\bm{a_2},2} \Big) + J_1 \Big(S^y_{-2\bm{a_1},2}+S^y_{-2\bm{a_2},2} \Big)   +J_z \Big(S^y_{(0, 0, -\frac{\bm{c}}{8}),3} \Big)\Biggr]  \Biggr\}.
 \label{eq:ds1x_dt_1}
\end{aligned}
\end{equation}
Substituting Eq.~\ref{eq:3_solutions}, we can rewrite this expression as
\begin{equation}
\begin{aligned}
  \frac{i\hbar\omega}{S}S^x_1&=-S^y_1\Bigl(2J+2J_1+J_z\Bigr)-S^y_3J_ze^{-i\frac{Q_z}{8}}\\
  &-S^y_2\Biggl[J\Bigl(e^{iQ_x}+e^{iQ_y}\Bigr) + J_1\Bigl(e^{-2iQ_x}+e^{-2iQ_y}\Bigr)\Biggr],
 \label{eq:ds1x_dt_2}
\end{aligned}
\end{equation}
and simplifying
\begin{equation}
 \frac{i\hbar\omega}{S}S^x_1=-AS^y_1-CS^y_2-DS^y_3,
 \label{eq:ds1x_dt_3}
\end{equation}
where we have defined
\begin{equation}
\begin{aligned}
  A&=2J+2J_1+J_z ,\\
  B&=A+J_z ,\\
  C&=J\left(e^{iQ_x}+e^{iQ_y}\right) + J_1\left(e^{-2iQ_x}+e^{-2iQ_y}\right), \\
  D&=J_ze^{-i\frac{Q_z}{8}}.
\label{eq:definitions}
\end{aligned}
\end{equation}
The final torque equations for the six spins are
\begin{equation}
\begin{aligned}
 \frac{i\hbar\omega}{S}S^x_1 &= -AS^y_1 -CS^y_2   -DS^y_3          ,  &\frac{i\hbar\omega}{S}S^y_1 &= +AS^x_1 +CS^x_2+ DS^x_3\\
 \frac{i\hbar\omega}{S}S^x_2 &= +AS^y_2 +C^*S^y_1 +DS^y_4          ,  &\frac{i\hbar\omega}{S}S^y_2 &= -AS^x_2 -C^*S^x_1 -DS^x_4\\
 \frac{i\hbar\omega}{S}S^x_3 &= +BS^y_3 +CS^y_4   +DS^y_5+D^*S^y_1 ,  &\frac{i\hbar\omega}{S}S^y_3 &= -BS^x_3 -CS^x_4 -DS^x_5-D^*S^x_1\\
 \frac{i\hbar\omega}{S}S^x_4 &= -BS^y_4 -C^*S^y_3 -DS^y_6-D^*S^y_2 ,  &\frac{i\hbar\omega}{S}S^y_4 &= +BS^x_4 +C^*S^x_3 +DS^x_6+D^*S^x_2\\
 \frac{i\hbar\omega}{S}S^x_5 &= -AS^y_5 -CS^y_6   -D^*S^y_3        ,  &\frac{i\hbar\omega}{S}S^y_5 &= +AS^x_5 +CS^x_6 +D^*S^x_3\\
 \frac{i\hbar\omega}{S}S^x_6 &= +AS^y_6 +C^*S^y_5 +D^*S^y_4        ,  &\frac{i\hbar\omega}{S}S^y_6 &= -AS^x_6 -C^*S^x_5 -D^*S^x_4 .
 \label{eq:full_torque}
&\end{aligned}
\end{equation}

\begin{figure}
\center
\includegraphics[width = 0.9\textwidth]{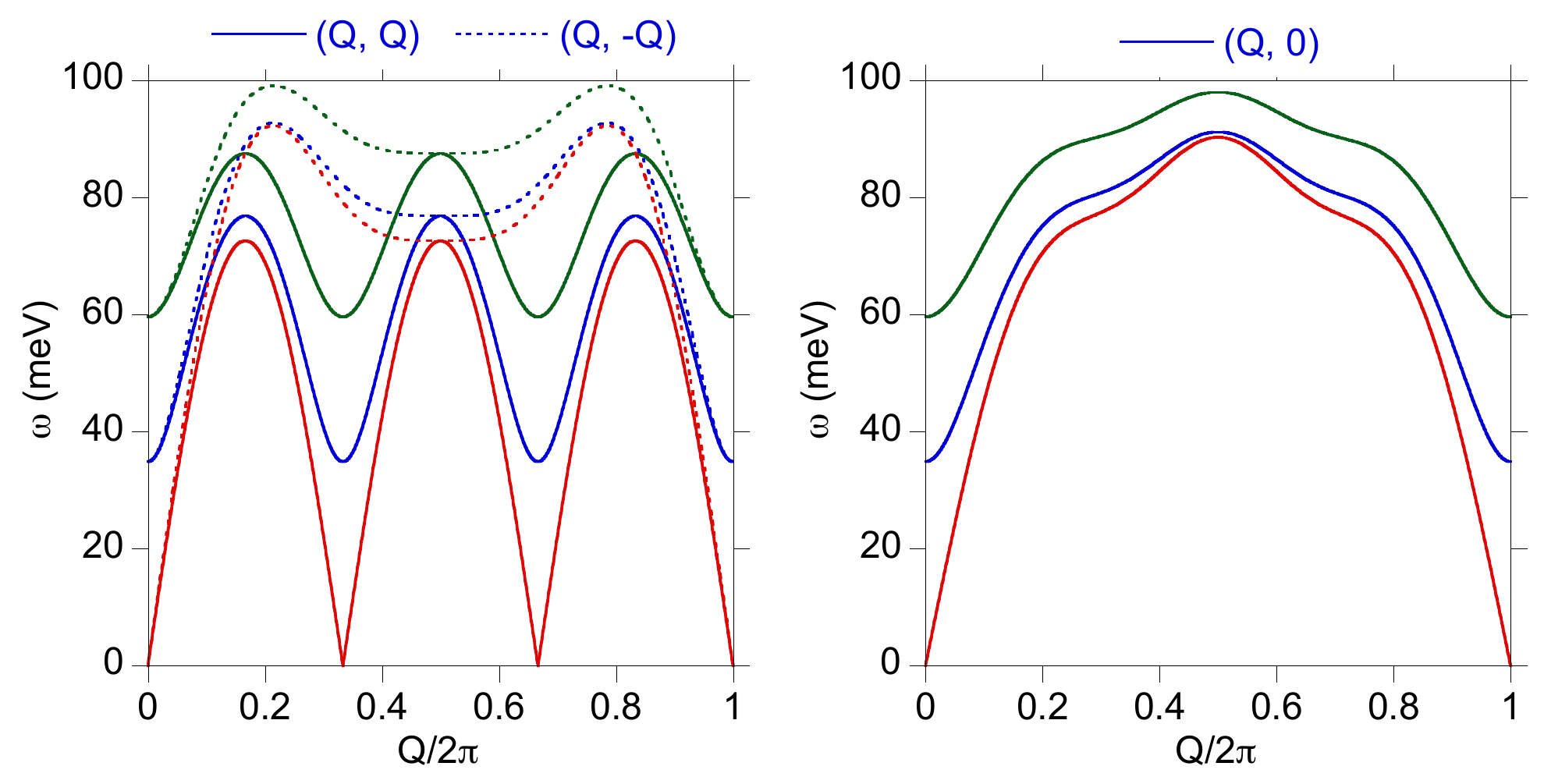}
\caption{Magnon dispersion with  $J$=69 meV, $J_1$=17 meV and $J_z$=13.6 meV, as in the main text, along the ($Q$,$Q$) and ($Q$,0) directions (with $Q_z$=0).  The dashed curves in the left plot are for the twin domain  ($Q$,-$Q$).}
\label{fig_magnon}
\end{figure}

This results in a $12\times12$ secular matrix (6 spins, 2 components, x, y, per spin)
\begin{equation}
 M=\begin{pmatrix}
    0    &    -A   &   0    &   -C    &    0    &  -D    &     0    &    0    &   0    &     0    &    0    &   0    \\ 
    A    &    0    &   C    &    0    &    D    &   0    &     0    &    0    &   0    &     0    &    0    &   0    \\
    0    &   C^*   &   0    &    A    &    0    &   0    &     0    &    D    &   0    &     0    &    0    &   0    \\ 
   -C^*  &    0    &  -A    &    0    &    0    &   0    &    -D    &    0    &   0    &     0    &    0    &   0    \\ 
    0    &   D^*   &   0    &    0    &    0    &   B    &     0    &    C    &   0    &     D    &    0    &   0    \\ 
   -D^*  &    0    &   0    &    0    &   -B    &   0    &    -C    &    0    &  -D    &     0    &    0    &   0    \\ 
    0    &    0    &   0    &   -D^*  &    0    &  -C^*  &     0    &   -B    &   0    &     0    &    0    &  -D    \\ 
    0    &    0    &  D^*   &    0    &   C^*   &   0    &     B    &    0    &   0    &     0    &    D    &   0    \\ 
    0    &    0    &   0    &    0    &    0    &  -D^*  &     0    &    0    &   0    &    -A    &    0    &  -C    \\ 
    0    &    0    &   0    &    0    &   D^*   &   0    &     0    &    0    &   A    &     0    &    C    &   0    \\ 
    0    &    0    &   0    &    0    &    0    &   0    &     0    &   D^*   &   0    &    C^*   &    0    &   A    \\ 
    0    &    0    &   0    &    0    &    0    &   0    &    -D^*  &    0    &  -C^*  &     0    &   -A    &   0    
\end{pmatrix}.
 \label{eq:Matrix}
\end{equation}
Diagonalizing this matrix yields the squared eigenvalues
\begin{equation}
  \lambda_1^2(\bm{Q})=-A^2+C^*C
   \label{eq:lambda1}
\end{equation}
and
\begin{equation}
  \lambda_{2,3}^2(\bm{Q})=-\frac{A^2}{2}-\frac{B^2}{2}+C^*C+2J^2_z\pm\frac{1}{2}J_z\sqrt{(A+B)^2+32C^*C-8J_z^2},
 \label{eq:lambda2}
\end{equation}
each of which are two-fold degenerate because of the tetragonal symmetry of the ground state.
As $i\omega=\lambda$, these eigenvalues correspond to the magnon branches
\begin{equation}
  \omega_\text{middle}(\bm{Q})=\sqrt{A^2-C^*C}
\label{eq:disp_rel1}
\end{equation} 
and
\begin{equation}
  \omega_\text{acoustic,optic}(\bm{Q})=\sqrt{\frac{A^2}{2}+\frac{B^2}{2}-C^*C-2J^2_z\mp\frac{1}{2}J_z\sqrt{(A+B)^2+32C^*C-8J_z^2}}. 
 \label{eq:disp_rel2}
\end{equation}
To justify these labels, and give some sense of these energies, we consider the $\Gamma$ point. Substituting $\bm{Q}=0$ in Eq.~\ref{eq:definitions} gives $C=2J+2J_1$ and $D=J_z$, which shows that our labels were chosen in order of increasing energy:
\begin{equation}
 \begin{aligned}
  &\omega_{\text{acoustic}}(\bm{0})=0, \\
  &\omega_{\text{middle}}(\bm{0})=SJ_z\sqrt{1+2C/J_z}\sim S\sqrt{2J_zC}, \\
  &\omega_{\text{optic}}(\bm{0})=SJ_z\sqrt{1+6C/J_z}\sim S\sqrt{6J_zC}, \\
 \end{aligned}
 \label{eq:disp_gamma}
\end{equation}
where the approximation applies for $J_z \ll J$.  The magnon dispersion for the parameters determined by the fit to the RIXS data, $J$=69 meV, $J_1$=17 meV (with $J_z$=13.6 meV), can be seen in Fig.~\ref{fig_magnon}.

\subsection{Calculation of the RIXS intensities}
As explained in the main text and Ref.~\cite{Haverkort2010theory}, the RIXS intensity for each mode, labeled $n$, can be written as
\begin{equation}
I_n(\bm{Q})=\left\lvert\sum_i \bm{k}_{in}\cdot\bm{M}_{n,\bm{Q}}(\bm{r}_i)\right\rvert^2,
 \label{eq:intensity_1}
\end{equation}
with $i$ summed over the six spins in the magnetic unit cell and
\begin{equation}
 \bm{M}_{n,\bm{Q}}(\bm{r}_i)=\left(c^x_{n,\bm{Q},i}S^x_i ,c^y_{n,\bm{Q},i}S^y_i, 0 \right).
 \label{eq:eigenvector_form}
\end{equation}
As the analytic expressions for the eigenvectors are complicated, we chose to determine them by diagonalizing the secular matrix Eq.~\ref{eq:Matrix} numerically for each $\bm{Q}$ using SciPy \cite{SciPy}. As noted above, each distinct magnon branch within our model has two degenerate eigenvalues, with the two members of each pair related by a 90 degree in-plane rotation because of the tetragonal symmetry. That is
\begin{equation}
\left(c_1^x,c_1^y,c_2^x,c_2^y,...\right)
\end{equation}
is degenerate with
\begin{equation}
\left(-c_1^y,c_1^x,-c_2^y,c_2^x,...\right) \label{eq:eivecs_ortho}
\end{equation}
with $n,\bm{Q}$ being implicit.  It is important to enforce this symmetry when calculating the RIXS intensity.

For the fit shown in the text, we took into account the scattering geometry of the RIXS measurements, with
\begin{equation}
 \bm{k}_{in}=\cos\theta\bm{I}-\sin\theta\bm{Q},
 \label{eq:kin_1}
\end{equation}
and
\begin{equation}
 \abs{\bm{Q}}=\frac{4\pi\sin\theta}{\lambda_E}
 \label{eq:mod_Q}
\end{equation}
with $\theta$ the Bragg angle and $\lambda_E$ the photon wavelength.  Here,
\begin{equation}
 \bm{I}=\frac{\bm{Q}\times\bm{c}\times\bm{Q}}{\abs{\bm{c}\times\bm{Q}}}.
 \label{eq:vector_I}
\end{equation}
In particular, as the in-plane component of $\bm{Q}$ is swept, the $Q_z$ component changes accordingly.

\section{First-principles determination of exchange coupling constants in \texorpdfstring{L\tlc{a}$_4$N\tlc{i}$_3$O$_8$}{La4Ni3O8}}

\begin{figure*}
\center
\includegraphics[width=\textwidth]{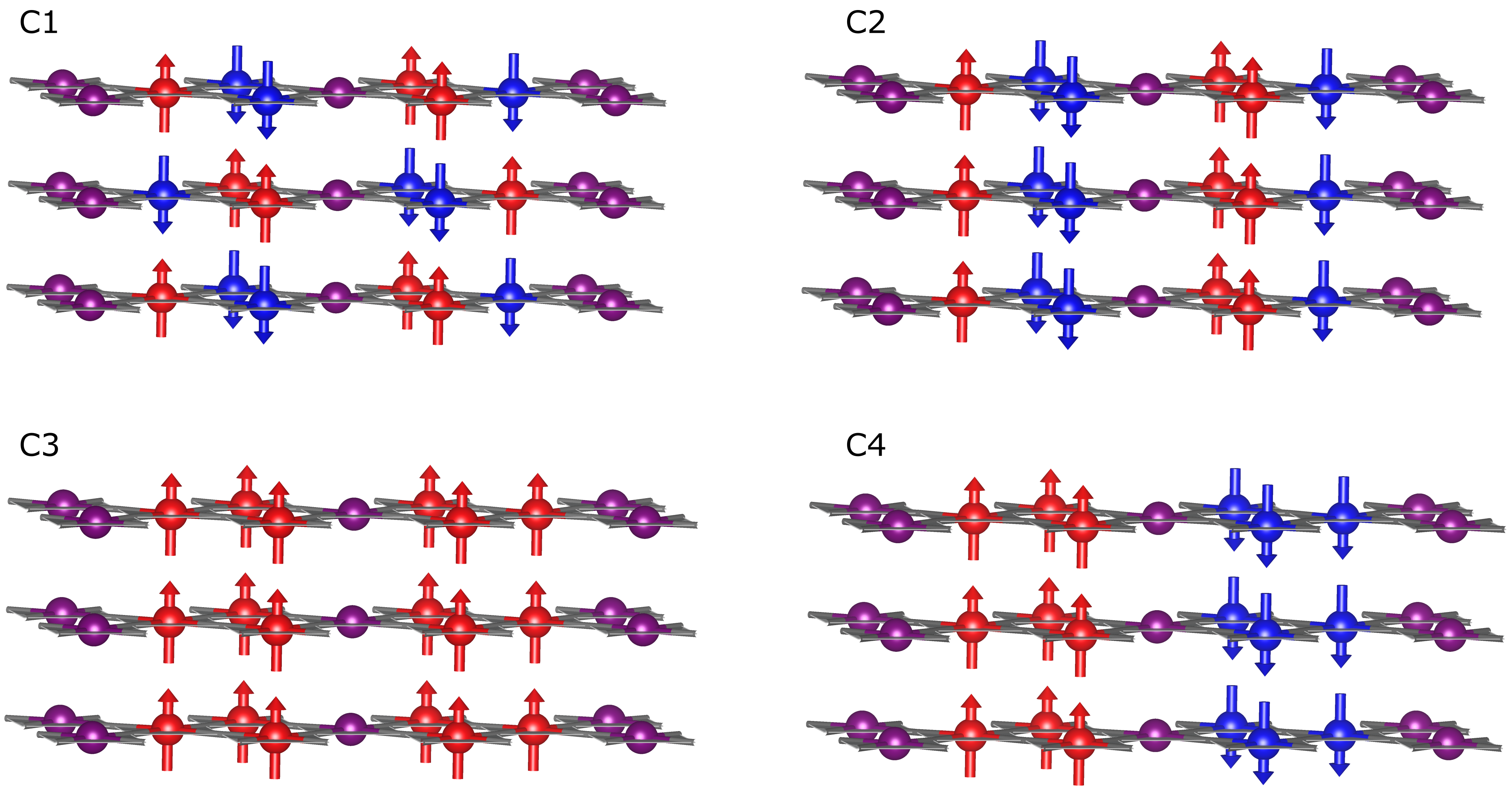}
\caption{Magnetic supercells used for determining the magnetic exchange. Ni sites are shown as spheres with red, blue and purple denoting up-spin, down-spin and hole states, respectively, as in Fig.~1 of the main text and Fig.~\ref{fig_couplings}. Gray lines trace the Ni-O bond network. Each magnetic cell is a $3\sqrt{2}a \times \sqrt{2}a \times c$ tiling of the structural cell  containing 36 Ni atoms (180 atoms in total), noting that there are two
trilayer units a given unit cell.  Atoms that are at the edge of the unit cell are shown as full spheres rather than being cut-off.}
\label{fig_exchange_coupling}
\end{figure*}

We performed \gls*{DFT} calculations for \LNO{}  with the all-electron full potential code WIEN2k \cite{Blaha2020WIEN2k1, Blaha2020WIEN2k2} using the \gls*{GGA} exchange-correlation functional \cite{Perdew1996generalized}. In these calculations, as we did previously \cite{Botana2016charge, Zhang2017large}, we considered the influence of thee Coulomb interaction, $U$. The $U$ modification of \gls*{DFT} is usually included to compensate for the under-localization of transition-metal $3d$-electrons in \gls*{DFT}. Including $U$, conversely, tends to over-localize electrons as it ``double counts'' the true Coulomb and exchange interactions as explained in, for example, Refs.~\cite{Ylvisaker2009anisotropy, Chen2016spin, Ryee2018effect}. Because of this, it is not immediately obvious whether including $U$ leads to a more accurate value of $J$. In calculations, we found that the inclusion of $U$ simply increases the size of the gap, and also leads to a modest increase in $J$, as outlined in Table~\ref{tab_DFT_J}. Calculations with and without $U$ find the same insulating, charge and spin stripe-ordered ground state, which was predicted before this state was experimentally observed \cite{Zhang2019stripe}. In our prior work, we examined the role of $U$ at length and found that the low-spin stripe state \LNO{} was appropriately described by \gls*{GGA} calculations even without the inclusion of $U$ \cite{Botana2016charge, Zhang2017large}. We refer the reader to these papers for the reasoning behind this. Since it allows us to compute $J$ with fewer adjustable parameters, we report values from \gls*{GGA} calculations in the main text, and find that this is in good accord with experiment. As is evident based on how $J$ changes with $U$, the very close match between theory and experiment of 2~meV  at the \gls*{GGA} level is likely coincidental. We note that similar values for the superexchange can be obtained from rough estimates using a sum of the Mott and charge transfer contributions, $J = 2t^2/U_d+2t^2/[\Delta+U_p/2]$, with $t= t_{pd}^2/\Delta$.
Using $U_d=8.5$~eV and $U_p=7.3$~eV values from a similar analysis on the cuprates \cite{McMahan1990}, with a $\Delta$ and $t_{pd}$ obtained from our Wannier fit for \LNO{} \cite{Nica2020theoretical}, a comparable $J$ of 99~meV is obtained.  But this, obviously, depends on the choice of $U_d$ and $U_p$. Spin-orbit coupling is not expected to have an appreciable effect on the exchange constants for $3d$ transition metal ions, especially for $e_g$ states where the orbital moment is largely quenched. This has been explicitly verified in our prior calculations \cite{Zhang2019stripe}.

The exchange couplings ($J$, $J_1$ and $J_z$) were obtained from total energy calculations for different Ni spin configurations (labeled C1-C4 in Fig.~\ref{fig_exchange_coupling}) mapped to a Heisenberg model. Configuration C1 is the experimental and theoretical ground state as shown in Fig.~1 of the main text. The magnitudes of the magnetic moments of the Ni$^{2+}$ atoms were between 0.6-0.7~$\mu_B$ and we confirmed that these values were similar within 0.1~$\mu_B$ in the different configurations, an accuracy typical of this type of calculation, justifying the Heisenberg mapping.  Different configurations C1-C4 have differing magnetic bonds:
\begin{itemize}
    \item C1 -- AFM $J$, AFM $J_1$, AFM $J_z$
    \item C2 -- AFM $J$, AFM $J_1$, FM $J_z$
    \item C3 -- FM $J$, FM $J_1$, FM $J_z$
    \item C4 -- FM $J$, AFM $J_1$, FM $J_z$
\end{itemize}
 The energies per trilayer are (with each bond counted once)
\begin{equation}
\begin{aligned}
E_{\text{C1}} &= E_0 - 4\times3 J S^2 - 3 \times 4 J_1 S^2 - 2\times 4 J_z S^2\\
              &= E_0 -3J - 3J_1 - 2 J_z\\
E_{\text{C2}} &= E_0 -3J - 3J_1 + 2 J_z \\
E_{\text{C3}} &= E_0 +3J + 3J_1 + 2 J_z\\
E_{\text{C4}} &= E_0 +3J - 3J_1 + 2 J_z
 \end{aligned}
 \label{eq:energies_spins}
\end{equation}
where $E_0$ is the non-magnetic energy. Solving this set of linear equations gives
\begin{equation}
\begin{aligned}
J   &= (E_{\text{C4}} - E_{\text{C2}})/6\\
J_1 &= (E_{\text{C3}} - E_{\text{C4}})/6\\
J_z &= (E_{\text{C2}} - E_{\text{C1}})/4.
 \end{aligned}
 \label{eq:couplings_energies}
\end{equation}
whose values are listed in the main text.

\begin{table}[b]
\begin{ruledtabular}
\caption{Exchange constants in the GGA and GGA+$U$ approximations ($U= 4.75$~eV).}
\begin{tabular}{ccc|ccc}
\label{tab_DFT_J}
Exchange constant & GGA+U (meV) & GGA (meV) &
Exchange ratio & GGA+U & GGA\\
\hline
$J$               & 97.60       & 71.40     & $J/J_z$           & 5.04        & 5.25  \\
$J_1$             & 14.28       & 10.65     & $J/J_1$           & 6.83        & 6.7    \\
$J_z$             & 19.38       & 13.6      & $J_z/J_1$         & 1.36        & 1.28   \\
\end{tabular}
\end{ruledtabular}
\end{table}

\section{X-Ray Absorption}
Our \gls*{XAS} measurements aim to compare the intensity of the O $K$-edge pre-peak feature as a guide to the $3d$-$2p$ orbital hybridization in nickelates and cuprates. We analyze our data on La$_4$Ni$_3$O$_8$ against literature data for La$_{2-x}$Sr$_x$CuO$_4$ from Ref.~\cite{Chen1992out} and Nd$_{1-x}$Sr$_x$NiO$_2$ from Ref.~\cite{Goodge2020doping}. As far as we are aware, Nd$_{1-x}$Sr$_x$NiO$_2$ data are only available for in-plane polarization, so we only show this component of the polarization. The intensity of the spectra are scaled to have equivalent intensities for energies above 538~eV past the main O $K$-edge step. Since different measurements can have different absolute energy calibrations, we used measurements of different reference samples to put the spectra on the same energy-scale. We use Ref.~\cite{Zhang2017large} as our reference energy calibration for which La$_4$Ni$_3$O$_8$ was measured alongside La$_{2-x}$Sr$_x$CuO$_4$ and SrTiO$_3$. We then used the SrTiO$_3$ reference measurements in \cite{Goodge2020doping} to put Nd$_{1-x}$Sr$_x$NiO$_2$ on the same energy scale. While in La$_4$Ni$_3$O$_8$ and La$_{2-x}$Sr$_x$CuO$_4$ the pre-peak is very clear, the pre-peak in Nd$_{1-x}$Sr$_x$NiO$_2$ is broader making isolating the pre-preak less immediately obvious. We took the same `background' intensity as was used in \cite{Goodge2020doping}, which comes from a measurement of undoped NdNiO$_2$ and use a lorentzian lineshape to fit. The dominant error in our analysis likely comes from inhomogeneity in the doping of the Nd$_{1-x}$Sr$_x$NiO$_2$ results we compare to and some uncertainty in how to isolate all the intensity in the pre-peak. These will likely improve in the future by higher quality sample preparation.  Further analysis of the ligand-hole anisotropy will also be important, but also requires polarization-dependent measurements of Nd$_{1-x}$Sr$_x$NiO$_2$ that are not currently available in the literature.
\bibliography{refs}